# K-method of cognitive mapping analysis

A.A.Snarskii[a,b], D.I.Zorinets[a,1], D.V.Lande[a,b], A.V.Levchenko[c]

[a] *NTUU "Kyiv Polytechnic Institute", Kiev, Ukraine*

[b] *Institute for Information Recording NASU, Kiev, Ukraine*

[c] *ROSEN Europe B.V., Kiev, Ukraine*

**Abstract**

Introduced a new calculation method (K-method) for cognitive maps. K - method consists of two consecutive steps. In the first stage, allocated subgraph composed of all paths from one selected node (concept) to another node (concept) from the cognitive map (directed weighted graph) . In the second stage, after the transition to an undirected graph (symmetrization adjacency matrix) the influence of one node to another calculated with Kirchhoff method. In the proposed method, there is no problem inherent in the impulse method. In addition to "pair" influence of one node to another, the average characteristics are introduced, allowing to calculate the impact of the selected node to all other nodes and the influence of all on the one selected. For impulse method similar to the average characteristics in the case where the pulse method "works" are introduced and compared with the K-method

## 1. INTRODUCTION

One of the directions in modeling of complicated network structures is the creation of cognitive maps, their description and analysis [1]. A cognitive map is an orientated graph to the edges of which weights may be collated. As any graph a cognitive map is defined by adjacency matrix, the elements of which correspond to the weights of the edges between the vertices. Some concepts correspond to the vertices of a cognitive map and cause-effect (casual) relations between concepts correspond to the edges (relations). The weight of the cognitive map edge has a positive value if the increasing of the weight of the concept-cause leads to the increasing of the weight of the concept. The negative weight of the edge means that the increasing of the weight of the concept-cause leads to the decreasing of the weight of the concept-effect. Thus, a cognitive map is an digraph to each concept of which the weight is assigned and some concept corresponds to each concept.

Fundamental models of cognitive maps are sign digraphs, weight digraphs and so-called fuzzy cognitive maps. The peculiarity of cognitive maps, being investigated now, is their weak structuralization which is so characteristic to social, economic, organizing, political and many other networks. In this paper weight digraphs are regarded as cognitive maps, and sign digraphs are interpreted as weight ones with weight value $\pm 1$.

Thus, a cognitive map is the digraph to each relation of which weight is assigned and concept of a cognitive map corresponds to each concept

In the first point the most frequently utilized method of quantitative description of cognitive maps –impulse method and the principal drawbacks so characteristic to it are very briefly considered. In the second point a new method of quantitative description of cognitive maps is described, we call it K-method for short. In point 3 some examples of calculations of cognitive maps with the help of K-method are given. In point 4 two characteristics of the obtained K-matrix, analogs of characteristics in HITS – algorithm proposed in [2], which are called "pressure" and "consequence" have been introduced and analyzed. In point 5 the cognitive maps in which the series of adjacency matrix diverge have been considered as a result this map cannot be investigated in the frames of impulse method. In the supplement for method characteristics which are analogous to the introduced for K-method "pressure" and "consequence" have been represented. The comparison of these characteristics has been made in case when the calculation of cognitive map in pulse method is possible.

---

[1] Corresponding author at: Kyiv Polytechnic Institute, Kyiv, Ukraine. D.I. Zorinets. deniszorinets@gmail.com

## 2. THE IMPULSE METHOD

According to the most frequently utilized method of quantitative description of cognitive maps - the impulse method, to each concept $i$ – some primordial value $v_i$(initial) is assigned. The problem is to define the final value of concept $v_i(t \to \infty)$, or in some cases the rate of change in time.

For definition $v_i(t)$ it is necessary to set the law of value change of the concept depending on its initial value, values of related neighboring to it concepts, weights of relations.

The basic procedure of cognitive mapping analysis is iterational method, described in [1] in detail. According to this method the values of each concept $v_i(n)$ at the moment of discrete time $n$ ($n = 1,2 \dots$) is defined by the following rule

$$\mathbf{v}(n+1) = \mathbf{v}(n) + \mathbf{W}\mathbf{p}(n), \quad n = 0,1,\dots \quad , \tag{1}$$

where $\mathbf{v}(n)$ –vector-column of net concepts of cognitive map $\mathbf{W}$ – adjacency matrix of digraph of relations with weight, and

$$\mathbf{p}(n) = \mathbf{v}(n) - \mathbf{v}(n-1), n = 1,2, \dots \quad . \tag{2}$$

At the initial moment of time it is considered set $\mathbf{p}(0)$ and $\mathbf{v}(init.)$ and

$$\mathbf{v}(0) = \mathbf{v}(init.) + \mathbf{p}(0), \tag{3}$$

so that the equation for definition $\mathbf{v}(n)$ has the form [1]

$$\mathbf{v}(n) = \mathbf{v}(init.) + \sum_{k=0}^{n} \mathbf{W}^k \mathbf{p}(0). \tag{4}$$

In the case when matrix series in (4) converges, $\mathbf{v}(n \to \infty)$ may be expressed through the matrix opposite to $\mathbf{1} - \mathbf{W}$, where $\mathbf{1}$ – unit matrix

$$\mathbf{v}(\infty) = \mathbf{v}(init.) + (\mathbf{1} - \mathbf{W})^{-1}\mathbf{p}(0), \tag{5}$$

further, there where it will not cause misunderstanding, we shall mark $\mathbf{v}(\infty)$ as $\mathbf{v}$.

While investigating of cognitive maps, initial values $\mathbf{v}(init.)$ on digraph concepts and pulse values $\mathbf{p}(0)$ at zero moment of time are set.

Investigation of cognitive maps according to (3) and (4), and when the series in (4) converges so and (5) is widely used in various situations [1, 3, 4]. It is necessary to note that in that approach to the cognitive mapping analysis there are a number of problems and contradictions, in some cases they do not allow to solve the problem of investigating of cognitive map. Let us enumerate some of them:

1. Diverging $\mathbf{v}(n)$ with $n \to \infty$, in $\infty$ step, when series in (4) diverge.
2. The result of calculation - $\mathbf{v}(n)$ depends according to (4) on the initial values $\mathbf{p}(0)$.
3. Initial value $\mathbf{v}(init.)$ does not at all affect the dependence $\mathbf{v}(n)$ from $n$ (comes into the expression for $\mathbf{v}(n)$ *as item*).
4. Increasing of elements of matrix $\mathbf{W}$ in the same value, not only changes the value component of vector $\mathbf{v}(n)$, but changes their rank distribution.

Thus, if series in (4) does not converge (that graph is called unstable),so there is no possibility to find $\mathbf{v}(\infty)$ and is left only to follow the changing $\mathbf{v}(n)$ with the increasing $n$,moreover, the numerical value of the vector components $\mathbf{v}(n)$, as a rule, quickly reach the values, going out of rational limits. There exist methods of stabilization of unstable graphs. Adjacency matrix $\mathbf{W}$ may be standardized [3], introduce or delete graph relations [2].The result of all these methods is that the adjacency matrix $\mathbf{W}$ of the cognitive map is substituted for another one. Beforehand it is not clear how the results obtained for the changed matrix $\mathbf{W}$ correlate with the initial problem.

F. Roberts in [3] gave an obvious example for the simplest cognitive map which consisted of two concepts and two relations for two variants from weights which show the profundity of the problem. In the first variant the weights were equal to $2$ and $-2$, but in the second $-1/2$ and $-1/2$. In the first variant series in (4) converges already with $n = 10$, in the second – diverges, reaching with the unit initial vector $\mathbf{p}(0)$, the values of order $10^{30}$ already for the hundredth step according to the time ($n = 100$).In addition, with the increasing of $n$, values $v_1(n)$ and $v_2(n)$, where the components of vector $\mathbf{v}(n)$ chaotically change from positive values to negative ones. It is quite clear that the considered variants differ one from the other only" by the choice of concepts of measure" and it must not affect the final result.

That is, at least, in some cases pulsed method [1, 3] has got principal drawbacks. In connection with this the urgent task is to work out new methods of investigation of cognitive maps. In this paper we propose $K$ –method in which there is no problem of series converging.

### 3. К –METHOD FOR COGNITIVE MAPPING

For clarity let us explain $K$ – method with the example of the cognitive map given in [1, 3] – fig. 1.

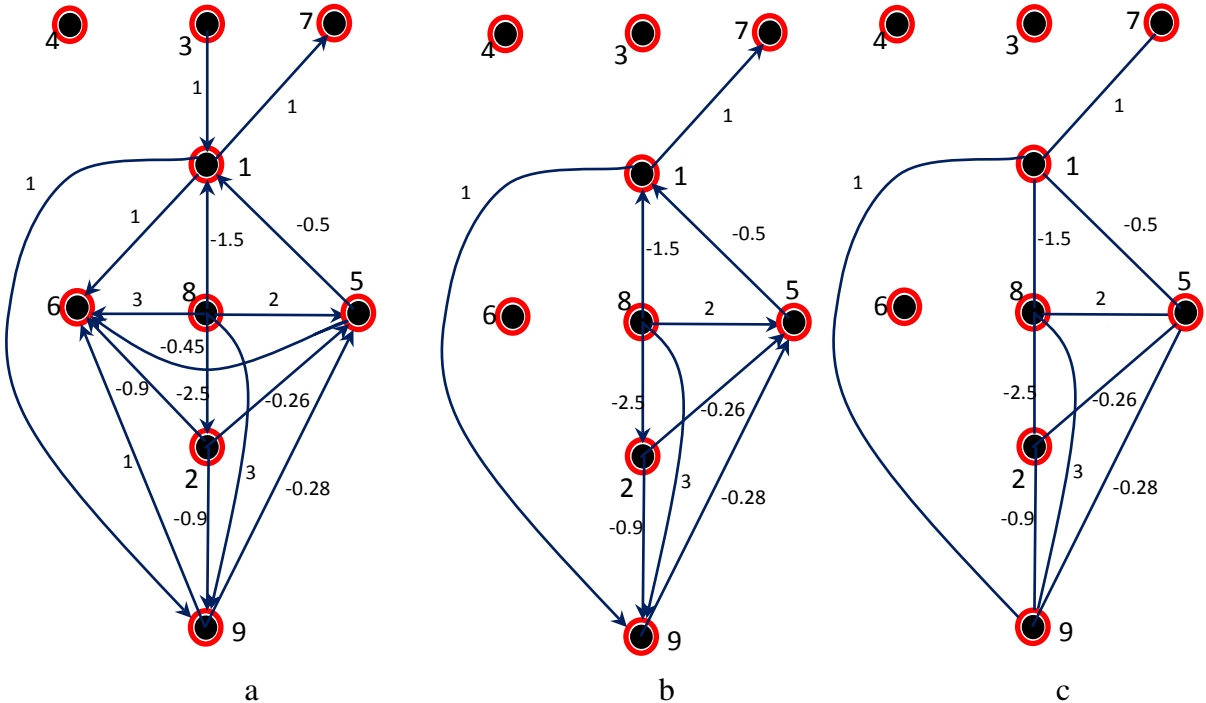

Fig. 1 Cognitive map. a – weighted graph for analysis of fuel consumption and maintenance of clean air in the city of San - Diego, California , where transportation is mainly made by cars. 1 – trip length; 2 – economy car fuel, mpg; 3 – population; 4 – cost of car; 5 – the cost of travel; 6 – emissions to the atmosphere; 7 – accidents; 8 – average delay; 9 – fuel consumption; b – highlighted subgraph $G(\alpha \to \beta)$, $\alpha = 1, \beta = 7$; c - symmetrized subgraph $B(\alpha \to \beta)$

Algorithm of $K$ – method consists of two stages.

Stage 1.For each directed pair of concepts $\alpha \to \beta$ we select subgraph $G(\alpha \to \beta)$ which consists of all possible ways (with the consideration of relation direction) from concept $\alpha$ to concept $\beta$. Thus, to each directed pair of concepts $\alpha \to \beta$ corresponds its own subgraph selected from complete cognitive map graph – **W**, let us mark the adjacency matrix of this subgraph and subgraph itself $G(\alpha \to \beta)$ – fig. 1b.

Stage 2. The stage of calculation of the directed influence of concepts on each other. Stage 2 consists of a number of steps. In the first step subgraph $G(\alpha \to \beta)$ symmetrizes and so symmetrized graph is marked as $B(\alpha \to \beta)$. All its relations become non-directed. So, for example, if in $\mathbf{G}(\alpha \to \beta)$ elements $\mathbf{G}_{km} \neq 0$, a $\mathbf{G}_{mk} = 0$, thus in symmetrized $\mathbf{B}(\alpha \to \beta)$ both elements are not equal to zero $\mathbf{B}_{km} \neq 0$, and $\mathbf{B}_{mk} = 0$ and $\mathbf{B}_{km} = \mathbf{B}_{mk} = \mathbf{G}_{km}$.

Adjacency matrix $\mathbf{B}(\alpha \to \beta)$ sets non-directed weight graph which further will be represented as net in which current of some preserved quantity, for example, electrical flows. With this analogy weights on relations are interpreted as electromotive force (emf), relations as resistances and quantities on concepts $v_i$ – as electrical potentials – $\varphi_i$ – fig. 1c.

At the second step calculation $\varphi_i$ for all concepts is done according to Kirchhoff rules [ ]. Since potentials are calculated with the accuracy to arbitrary constant, only their difference has got sense. For definiteness it is easy to set it on one of the concepts equal to zero. For subgraph $B(\alpha \to \beta)$ potential will be counted off from concepts $\alpha$, that is $\varphi_\alpha = 0$ will be accepted.

Equation for the calculation of potentials for all the rest concepts of net $B(\alpha \to \beta)$, according to Kirchhoff rules [5], that can be written in form

$$\boldsymbol{\Omega}(\alpha \to \beta)\mathbf{Y}\boldsymbol{\Omega}^T(\alpha \to \beta)\boldsymbol{\varphi}(\alpha \to \beta) = -\boldsymbol{\Omega}(\alpha \to \beta)\mathbf{Y}\mathbf{E}. \tag{6}$$

Here $\boldsymbol{\Omega}(\alpha \to \beta)$ – incident matrix for subgraph $B(\alpha \to \beta)$. This matrix is composed in the following way. Number is conferred to each relation and concept. The number of matrix row $\boldsymbol{\Omega}(\alpha \to \beta)$ corresponds to the number of relation, and the number of column – the number of concept. Element $(i,j)$ matrix $\boldsymbol{\Omega}(\alpha \to \beta)$ - $\omega_{ij}$ equals $+1$, if the relation with number $i$ goes out from concept $j$ and $-1$ – if comes into it. If relation $i$ is not connected with concept $j$, then $\omega_{ij} = 0$. In addition, since it is accepted $\varphi_\alpha = 0$, $\alpha$ the column of matrix $\boldsymbol{\Omega}(\alpha \to \beta)$ is deleted. Matrix $\mathbf{Y}$ – square matrix of resistances of dimension $M \times M$, where $M$ – the number of relations, moreover, each element of the given matrix - $y_{ij}$ is defined as

$$y_{ij} = R_i \delta_{ij}, \tag{7}$$

where $R_i$ – resistance of that relation and, $\delta_{ij}$ – Kronecker symbol.

In this paper all relations are accepted with resistances equal one and so $\mathbf{Y} \equiv \mathbf{1}$. In this case $\boldsymbol{\Omega}(\alpha \to \beta)\mathbf{Y}\boldsymbol{\Omega}^T(\alpha \to \beta) = \boldsymbol{\Omega}(\alpha \to \beta)\boldsymbol{\Omega}^T(\alpha \to \beta)$, and (6) gets the following form

$$\boldsymbol{\Omega}(\alpha \to \beta)\boldsymbol{\Omega}^T(\alpha \to \beta)\boldsymbol{\varphi}(\alpha \to \beta) = -\boldsymbol{\Omega}(\alpha \to \beta)\mathbf{E}. \tag{8}$$

Vector $\mathbf{E}$ -vector column consisting of $M$ elements, where $M$ – the number of relations and its element $e_i$ is defined as

$$e_i = \varepsilon_i, \tag{9}$$

where $e_i$ – emf on the relation with number $i$.

Selecting from initial cognitive map subgraph $G(\alpha \to \beta)$ and corresponding to it subgraph $B(\alpha \to \beta)$ we take into account that we are studying the influence of initial concept $\alpha$ on concept $\beta$, and that with the given (electrical) analogy means the difference of values $\boldsymbol{\varphi}_\beta(\alpha \to \beta) - \boldsymbol{\varphi}_\alpha(\alpha \to u\beta)$ and since accepted that $\boldsymbol{\varphi}_\alpha(\alpha \to \beta) = 0$, value $\boldsymbol{\varphi}_\beta(\alpha \to \beta)$. That is from the whole vector $\boldsymbol{\varphi}(\alpha \to \beta)$, which is set by solving equation (6)

$$\boldsymbol{\varphi}(\alpha \to \beta) = -[\boldsymbol{\Omega}(\alpha \to \beta)\mathbf{Y}\boldsymbol{\Omega}^T(\alpha \to \beta)]^{-1}\boldsymbol{\Omega}(\alpha \to \beta)\mathbf{Y}\mathbf{E}, \tag{10}$$

For further analysis only one component $\boldsymbol{\varphi}_\beta(\alpha \to \beta)$ is left.

With the same fixed $\alpha$, giving different values $\beta$ (and after all calculations) we find a new vector $\boldsymbol{\Phi}_\beta(\alpha \to \beta), \beta = 1,2, \ldots, N, \beta \neq \alpha$. It is necessary to note that if the components of vector $\boldsymbol{\varphi}(\alpha \to \beta)$ are related to the same graph set by matrix $\mathbf{B}(\alpha \to \beta)$ $(\boldsymbol{\varphi}_1(\alpha \to \beta), \boldsymbol{\varphi}_2(\alpha \to \beta), \ldots, \boldsymbol{\varphi}_N(\alpha \to \beta))$, then each component of vector $\boldsymbol{\Phi}(\alpha \to \beta)$ is related to its graph $\Phi(\alpha \to \beta)$ $(\boldsymbol{\Phi}_1(\alpha \to \beta), \boldsymbol{\Phi}_2(\alpha \to \beta), \ldots, \boldsymbol{\Phi}_N(\alpha \to \beta))$.

The set of all components of all vectors $\boldsymbol{\Phi}(\alpha \to \beta)$ may be written down in the form of matrix, we may call it $K$– matrix. To calculate element $K_{\alpha\beta}$ of matrix $\mathbf{K}$, we choose two concepts $\alpha \to \beta$ in cognitive map, select subgraph $G(\alpha \to \beta)$ of all possible relations out of concepts $\alpha$ in concept $\beta$, graph $G(\alpha \to \beta)$, transform in graph $B(\alpha \to \beta)$, attribute zero potential to concept $\alpha$ and according to (10) calculate the potential of concept $\beta$, the value of which is element $K_{\alpha\beta}$. That is, to each element of matrix $\mathbf{K}$ subgraph of cognitive map corresponds.

## 4. SOME SIMPLE EXAMPLES

To explain how $K$ – method works let us compare it with impulse - method and consider two simple cognitive maps. The first map marked $Q$ – fig. 2a consists of two concepts and one relation and the second one marked – $S$ consists of two concepts and two relations – fig. 2b.

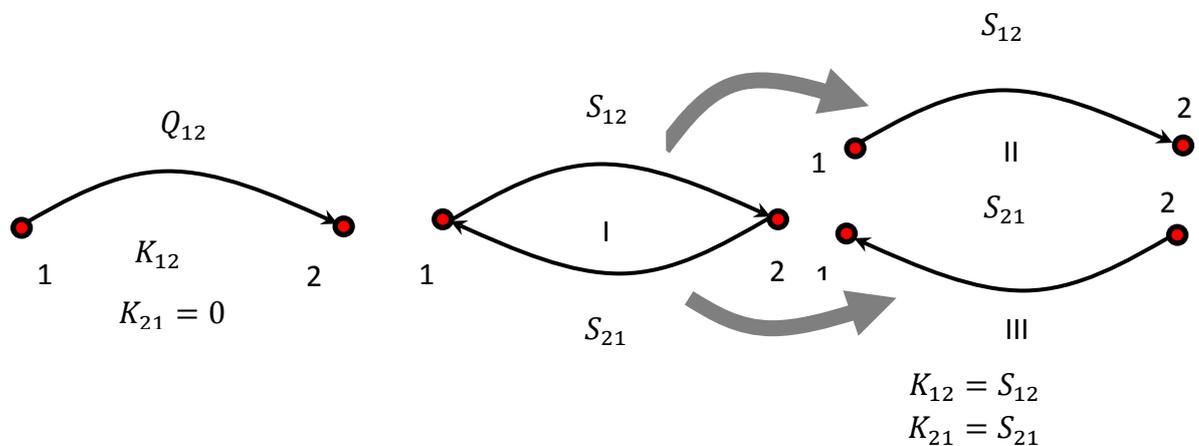

*Fig.* 2 Two examples of the simplest cognitive maps

In cognitive map $Q$ weight of relation setting the influence of concept 1 on concept 2 equals $Q_{12}$, concept 2 on concept 1 does not affect - $Q_{21} = 0$. In cognitive map $S$ weight of relation setting influence of concept 1 on concept 2 equals $S_{12}$, and concept 2 on concept 1 - $S_{21}$.

The calculation of matrix $\mathbf{K}$ for cognitive map $Q$ is simple, in this matrix $2 \times 2$ all elements equal zero, except one element $K_{12} = Q_{12}$.

In the case with cognitive map $S$ calculation is also simple. For calculating of influence of concept 1 on concept 2 from graph $S$ – fig. 2b-I it is necessary to select subgraph fig. 2b-II, from where at once follows $K_{12} = S_{12}$. Similarly for calculating of influence of concept 2 on concept 1 – fig. 2b-III $K_{21} = S_{21}$.

In [3] cognitive map $S$ (fig. 2b) has been considered with the help of iterational method for a number of cases. In the first case we have selected $S_{12} = S_{21} = 1/2$, that if we accept $\mathbf{v}(init.) = 0$ and $\mathbf{p}(0) = \mathbf{1}$, it gives $v_1 = v_2 = 2$. In $K$ – method with $S_{12} = S_{21} = 1/2$ influence of concept 1 on 2 and 2 on 1 equal $K_{12} = K_{21} = 1/2$. The second case which was considered in [3] differed from the first one only that both values $S_{12}$ and $S_{21}$ multiplied by 4 $S_{12} = S_{21} = 2$. That calculation shows that the influence of concepts on each other is multiplied by the same times - $v_1 = v_2 = 8$, $K_{12} = K_{21} = 2$. That is, both methods –impulse method and $K$ – method quantitatively coordinate with one another.

The following two cases considered in [3] are connected with the situation when weights have different signs, for example, $S_{12} = -1/2$ и $S_{21} = 1/2$. In this case К – method gives $K_{12} = -1/2$ and $K_{21} = 1/2$, that is, the influence of concepts is the same in module and opposite in sign, it corresponds to qualitative considerations.

Impulse method with $S_{12} = -S_{21} = 1/2$ also gives different signs and $v_1$ and $v_2$, although their modules are different- $v_1 = 0.8$, $v_2 = 0.4$, that with the symmetry of the problem seems strange.

The most obvious difference between К – method and pulsed method is observed if weights of relations in the previous example are multiplied by 4, that is if we choose $S_{12} = -S_{21} = 2$. With that choice series in (4) diverges and in pulse method already with $n = 100$ quantities $|v_1|$ и $|v_2|$ reach the values of order $10^{30}$. К – method with multiplying $S_{ik}$ by 4 leads to values $K_{ik}$ by 4 times larger.

Note that in pulsed method ratio $v_1(n)/v_2(n)$ remains final when $n$ increases and is a periodical function with period $n = 2$, accepting consecutively values $3, -1/3, 3, \dots$ .

## 5. MAIN CHARACTERISTICS OF COGNITIVE MAPS OBTAINED WITH $K$ – METHOD

Here we will show as an example some cognitive maps, which of their characteristics may be introduced and calculated with $K$ –method and compare these characteristics with the data obtained with impulse - method.

Consider the cognitive map proposed by F. Roberts [1, 3], see fig.1. Calculating for all $\alpha$ and $\beta$ of vector $\boldsymbol{\varphi}(\alpha \to \beta)$ according to (10) find $\mathbf{K}$ – matrix

$$\mathbf{K} = \begin{pmatrix} 0 & 0 & 0 & 0 & 0.72 & 0.94 & 1 & 0 & 1 \\ -1.28 & 0 & 0 & 0 & -0.573 & -0.835 & -0.133 & 0 & -0.295 \\ 1 & 0 & 0 & 0 & 1.72 & 1.457 & 2 & 0 & 2 \\ 0 & 0 & 0 & 0 & 0 & 0 & 0 & 0 & 0 \\ -0.5 & 0 & 0 & 0 & 0 & 0.038 & 0.5 & 0 & 0.5 \\ 0 & 0 & 0 & 0 & 0 & 0 & 0 & 0 & 0 \\ 0 & 0 & 0 & 0 & 0 & 0 & 0 & 0 & 0 \\ 0.256 & -2.5 & 0 & 0 & -0.702 & 1.388 & 2.482 & 0 & -1.205 \\ -0.78 & 0 & 0 & 0 & -0.28 & -3 \cdot 10^{-3} & 0.22 & 0 & 0 \end{pmatrix} \quad (11)$$

Ranking according to the quantity of $\mathbf{K}$ – matrix elements gives

$$K_{87} = 2.482, \quad K_{37} = 2, \quad K_{39} = 2, \dots, K_{89} = -1.205, K_{21} = -1.28, K_{82} = -2.5. \quad (12)$$

This ranked series of $\mathbf{K}$ elements coordinates with qualitative considerations. Really, according to (12) maximum influence $K_{87}$ exercises concept "mean delay" (concept 8) on concept "incidents" (concept 7). That is, the more the density of population the more wastes. And maximum negative influence ($K_{82}$) concept "mean delay" (concept 8) – concept "fuel economy" (concept 2).

It is necessary to note that for the concepts connected directly with each other, for example $(8 \to 6)$, value $K_{86} = 1.39$ does not coincide with the weight of relation $W_{86} = 3$. Of course, it means that in $K_{86}$ not only direct influence $W_{86} = 3$, comes but and mediate one, for example $W_{81}$, and then $W_{16}$ and so on.

Thus, according to (11) $K$ –method allows to calculate the influence of one concept (concept $\alpha$) on another one (concept $\beta$) with the account of all mediate influences through each other. We may call that kind of characteristic of cognitive map "pair interaction".

Besides that "pair interaction" on the base of $\mathbf{K}$ –matrix we may introduce and calculate so-called "collective interaction". Here we will introduce two characteristics which we will call pressure – $\boldsymbol{\psi}$ and consequence – $\boldsymbol{\nu}$, see fig. 3. It is necessary to note that the arrows on the relations in diagrams, fig. 3 do not mean relations in cognitive map but they mean the existence of corresponding matrix **K** component.

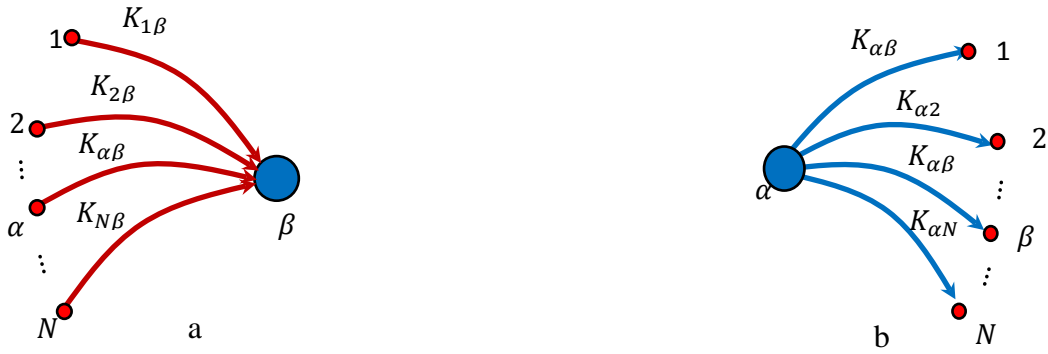

Fig. 3 Collective characteristics in $K$ – method and – pressure **ψ** , b -consequence **ν**

Components of vectors **ψ** and **ν** are connected with **K** –matrix components in the following way

$$\psi_\beta = \sum_\alpha K_{\alpha\beta}, \qquad \nu_\alpha = \sum_\beta K_{\alpha\beta}. \qquad (13)$$

Since in these cognitive maps we do not take into consideration the influence of the concept on itself items c $\alpha = \beta$ are not in sums (13).

According to its definition - (13) pressure $\psi_\beta$ describes the total influence of all other concepts on concept $\beta$ . And consequence $\nu_\alpha$ is the sum of influences of concept $\alpha$ on all other concepts.

Besides definition (13) we may introduce amplitude pressure **aψ** and consequence **aν**, for calculating of which sums in (13) are taken in absolute values

$$a\psi_\beta = \sum_\alpha |K_{\alpha\beta}|, \qquad a\nu_\alpha = \sum_\beta |K_{\alpha\beta}|. \qquad (14)$$

That is, for calculating, for example, $a\nu_\alpha$ the sign of concept $\alpha$ influence on concept $\beta$ is not important, we take into consideration only its value.

Calculation of pressures **ψ**, **aψ**, and consequences **ν**, **aν** for cognitive map represented in fig.1 are given in Table 1. Maximum value **ψ**, and **aψ** has the sixth concept "volume of waste in atmosphere' " $\psi_6 = 8.9$. This means that all concepts in sum influence the volume of wastes in atmosphere much more, increasing this volume. Minimum value both for **ψ** and for **aψ** occupies the same second concept – "cost of journey" $\psi_2 = -2.6$. This means that all other components in the system influence the cost of journey much more negatively, increasing it. The cost of journey "suffers" much more from the influence of other concepts.

| # | ψ | # | aψ | # | ν | # | aν |
|---|---|---|---|---|---|---|---|
| 7 | 6.609 | 7 | 6.335 | 3 | 8.117 | 8 | 8.533 |
| 6 | 2.985 | 9 | 5 | 1 | 3.66 | 3 | 8.177 |
| 9 | 3 | 6 | 4.661 | 5 | 0.538 | 1 | 3.66 |
| 5 | 0.885 | 5 | 3.995 | 4 | 0 | 5 | 3.116 |
| 3 | 0 | 1 | 3.816 | 6 | 0 | 2 | 1.158 |
| 4 | 0 | 2 | 2.5 | 7 | 0 | 9 | 1.283 |
| 8 | 0 | 3 | 0 | 8 | -0.281 | 4 | 0 |
| 1 | -1.304 | 4 | 0 | 9 | -0.843 | 6 | 0 |
| 2 | -2.5 | 8 | 0 | 2 | -3.116 | 7 | 0 |

Table 1. Ranked distribution of concepts according to values **ψ**, **aψ**, **ν**, and **aν** of cognitive map represented in fig.1

For characteristic which is called consequence – $\nu$, according to Table 1 ranked distribution gives the next. Maximum value for $\nu$, and $a\nu$ occupies the third concept – "fuel economy", it also coincides with qualitative considerations.

Note that in impulse method we may introduce a characteristic to some extent analogous to pressure $\psi$ and consequence $\nu$. In Appendix A the calculation of **K** matrix and $\psi^{imp}$, $a\psi^{imp}$, $\nu^{imp}$, and $a\nu^{imp}$ is given and their comparison with $\psi$, $a\psi$, $\nu$, and $a\nu$ is given.

In Appendix A the calculation of **K** – matrix and all above introduced characteristics for cognitive maps from [1] and from paper [4] is given. Comparison of some characteristics from these characteristics with the characteristics, obtained with pulsed method, is given.

## 6. COGNITIVE MAPS WITH DIVERGING SERIES

Here we will consider one more example of $K$ – method usage in the research of cognitive maps. Have a look at a cognitive map Cognitive map concerning some public health issues [4] – fig. 3. In an initial variant this cognitive map was with a sign – positive or negative sign was given to each relation fig 3a,

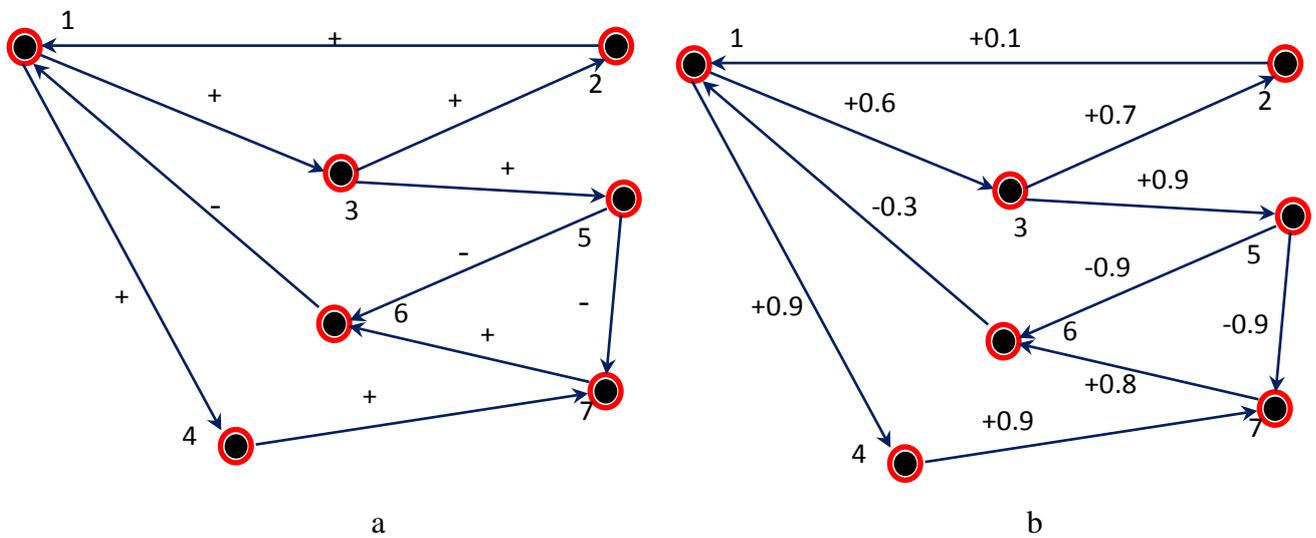

a
b

Fig. 4 a – with a sign Cognitive map concerning some public health issues. b – cognitive map concerning some public health issues. 1-Number of people in city, 2- Migration into city, 3 – Modernization, 4 – Garbage per area, 5 – Sanitation facilities, 6 – Number of diseases per 1000 residents, 7 – Bacteria per area

The simplest variant of investigation of that cognitive map with $K$ – method – introduce weight of relation when weight $+1$ corresponds to positive relation and $-1$-to negative one.

Adjacency matrix of that cognitive map has the form

$$\mathbf{W} = \begin{pmatrix} 0 & 0 & 1 & 1 & 0 & 0 & 0 \\ 1 & 0 & 0 & 0 & 0 & 0 & 0 \\ 0 & 1 & 0 & 0 & 1 & 0 & 0 \\ 0 & 0 & 0 & 0 & 0 & 0 & 1 \\ 0 & 0 & 0 & 0 & 0 & -1 & -1 \\ -1 & 0 & 0 & 0 & 0 & 0 & 0 \\ 0 & 0 & 0 & 0 & 0 & 1 & 0 \end{pmatrix}. \qquad (15)$$

**K** –matrix of the cognitive map with adjacency matrix **W** has the form

$$\mathbf{K} = \begin{pmatrix} 0 & 2 & 1 & 1 & 2 & 2 & 1.6 \\ 1 & 0 & 2 & 2 & 3 & 3 & 2.6 \\ 0.857 & 1 & 0 & 1.857 & 1 & 1.353 & 1.333 \\ 1 & 3 & 2 & 0 & 3 & 2 & 1 \\ -1.667 & 0.333 & -0.667 & -0.667 & 0 & -0.667 & -0.8 \\ -1 & 1 & 0 & 0 & 1 & 0 & 0.6 \\ 0 & 2 & 1 & 1 & 2 & 1 & 0 \end{pmatrix}. \quad (16)$$

According to (16) the first maximum elements of matrix **K** are $K_{25} = K_{26} = K_{42} = K_{45} = 3$, and minimum elements are - $K_{51} = -1.667$, $K_{61} = -1$, $K_{57} = -0.8$. This means that maximum influence has concept "Migration into city" on concept "Sanitation facilities".

Rank distributions of $a\psi$ and $av$ are given in table 2. It follows from the table that "Sanitation facilities" and "Number of diseases per 1000 residents" are the most vulnerable concepts and – "Migration into city" and "Garbage per area" are the most influential ones.

Comparison with impulse - method for cognitive map with adjacency matrix (15) is impossible, since maximum own value **W** $1.194 > 1$ and series (4) with $n \to \infty$ diverges. The simplest way to correct the situation is to standardize **W**, so that the series will converge. For example, if instead **W** we introduce $\mathbf{W_1} = \mathbf{W}/1.2$, then for standardized adjacency matrix $\mathbf{W_1}$ series (4) will converge.

For the investigation of cognitive map with that adjacency matrix impulse method may be used. The result of work is given in table 2.

| $\psi$ | # | # | $\psi^{imp}$ | $a\psi$ | # | # | $a\psi^{imp}$ |
|---|---|---|---|---|---|---|---|
| 12 | 5 | 4 | 143.079 | 12 | 5 | 1 | 307.351 |
| 9.333 | 2 | 1 | 90.397 | 10.02 | 6 | 4 | 274.205 |
| 8.686 | 6 | 2 | 69.57 | 9.333 | 2 | 3 | 223.709 |
| 6.333 | 7 | 5 | 53.841 | 7.933 | 7 | 2 | 220.232 |
| 5.333 | 3 | 3 | 42.914 | 6.667 | 3 | 5 | 204.503 |
| 5.19 | 4 | 7 | 19.868 | 6.524 | 4 | 6 | 156.689 |
| 0.19 | 1 | 6 | -90.397 | 5.524 | 1 | 7 | 35.033 |

| $v$ | # | # | $v^{imp}$ | $av$ | # | # | $av^{imp}$ |
|---|---|---|---|---|---|---|---|
| 13.6 | 2 | 3 | 257.848 | 13.6 | 2 | 3 | 347.583 |
| 12 | 4 | 5 | 193.974 | 12 | 4 | 5 | 260.265 |
| 9.6 | 1 | 1 | 138.477 | 9.6 | 1 | 1 | 188.146 |
| 7.4 | 3 | 2 | 128.51 | 7.4 | 3 | 6 | 179.007 |
| 7 | 7 | 4 | -83.576 | 7 | 7 | 2 | 169.901 |
| 1.6 | 6 | 7 | -129.954 | 4.801 | 5 | 7 | 163.113 |
| -4.135 | 5 | 6 | -179.007 | 3.6 | 6 | 4 | 113.709 |

Table 2. Rank distribution of concepts according to the values of components $\psi$, $a\psi$, $v$ and $av$ and $\psi^{imp}$, $a\psi^{imp}$, $v^{imp}$, and $av^{imp}$ for the cognitive map represented in fig. 1.

Note that rank distribution $\psi$, $a\psi$, $v$, and $av$, obtained with $K$ – method for cognitive map with adjacency matrix **W** and with matrix $\mathbf{W_1} = \mathbf{W}/1.2$ do not differ.

As you see from Table 2 for some concepts we notice a good correspondence, for example, according to rank the first two concepts $\psi$ and $\psi^{imp}$ coincide. But considerable distinctions also exist.

It is necessary to understand that from "the point of pulsed method view" standardization of adjacency matrix changes "the essence" of cognitive map. So if other standardization is used, for example, let us take $\mathbf{W_2} = \mathbf{W}/12$, rank distribution of characteristics obtained with pulsed method changes, see Table 3. As we see from Table 3 rank distribution for $\mathbf{W_1}$ and $\mathbf{W_2}$ is considerably different, as distinct from $K$ – method.

| $\mathbf{\psi}^{imp}(\mathbf{W_1})$ | # | | # | $\mathbf{\psi}^{imp}(\mathbf{W_2})$ | $\mathbf{a\psi}^{imp}(\mathbf{W_1})$ | # | | # | $\mathbf{a\psi}^{imp}(\mathbf{W_2})$ |
|---|---|---|---|---|---|---|---|---|---|
| 143 | 4 | | 5 | 0.09 | 307 | 1 | | 1 | 0.189 |
| 90 | 1 | | 2 | 0.09 | 274 | 4 | | 6 | 0.188 |
| 69 | 2 | | 4 | 0.084 | 224 | 3 | | 7 | 0.181 |
| 54 | 5 | | 3 | 0.083 | 220 | 2 | | 4 | 0.099 |
| 43 | 3 | | 1 | $8 \cdot 10^{-3}$ | 204 | 5 | | 3 | 0.099 |
| 20 | 7 | | 7 | $-5 \cdot 10^{-4}$ | 157 | 6 | | 5 | 0.092 |
| -90 | 6 | | 6 | $-8 \cdot 10^{-3}$ | 35 | 7 | | 2 | 0.091 |

| $\mathbf{v}^{imp}(\mathbf{W_1})$ | # | | # | $\mathbf{v}^{imp}(\mathbf{W_2})$ | $\mathbf{av}^{imp}(\mathbf{W_1})$ | # | | # | $\mathbf{av}^{imp}(\mathbf{W_2})$ |
|---|---|---|---|---|---|---|---|---|---|
| 258 | 3 | | 1 | 0.19 | 34 | 3 | | 3 | 0.189 |
| 194 | 5 | | 3 | 0.16 | 260 | 5 | | 1 | 0.187 |
| 138 | 1 | | 2 | 0.01 | 188 | 1 | | 5 | 0.182 |
| 128 | 2 | | 4 | 0.09 | 179 | 6 | | 6 | 0.099 |
| -83 | 4 | | 7 | 0.08 | 167 | 2 | | 2 | 0.098 |
| -130 | 7 | | 6 | -0.1 | 163 | 7 | | 7 | 0.092 |
| -179 | 6 | | 5 | -0.17 | 114 | 4 | | 4 | 0.091 |

Table 3. Rank distribution of concepts according to the values of components $\mathbf{\psi}^{imp}$, $\mathbf{a\psi}^{imp}$, $\mathbf{v}^{imp}$, and $\mathbf{av}^{imp}$ for cognitive maps with adjacency matrices $\mathbf{W_1}$ and $\mathbf{W_2}$.

In the same paper [4], in which the cognitive map represented in fig. 3 was given, the cognitive map with the same concepts but other weights is given, see. fig.3b. Its adjacency matrix has the form

$$\widetilde{\mathbf{W}} = \begin{pmatrix} 0 & 0 & 0.6 & 0.9 & 0 & 0 & 0 \\ 0.1 & 0 & 0 & 0 & 0 & 0 & 0 \\ 0 & 0.7 & 0 & 0 & 0.9 & 0 & 0 \\ 0 & 0 & 0 & 0 & 0 & 0 & 0.9 \\ 0 & 0 & 0 & 0 & 0 & -0.9 & -0.9 \\ -0.3 & 0 & 0 & 0 & 0 & 0 & 0 \\ 0 & 0 & 0 & 0 & 0 & 0.8 & 0 \end{pmatrix}. \qquad (17)$$

Module of maximum value $\widetilde{\mathbf{W}}$ is less than one ( $|-0.533 + 0.433i|$ =0.686) and the corresponding series in iterational method converges, that allows …$\mathbf{\psi}$ and $\mathbf{v}$ …with $K$ – method and iterational method.

$\mathbf{K}$ –matrix of cognitive map with adjacency matrix $\widetilde{\mathbf{W}}$ has the form

$$\mathbf{K} = \begin{pmatrix} 0 & 1.3 & 0.6 & 0.9 & 1.5 & 1.6 & 1.32 \\ 0.1 & 0 & 0.7 & 1 & 1.6 & 1.7 & 1.42 \\ 0.443 & 0.7 & 0 & 1.343 & 0.9 & 0.941 & 0.857 \\ 1.4 & 2.7 & 2 & 0 & 2.9 & 1.7 & 0.9 \\ -0.933 & 0.367 & -0.333 & -0.033 & 0 & -0.633 & -0.6 \\ -0.3 & 1 & 0.3 & 0.6 & 1.2 & 0 & 1.02 \\ 0.5 & 1.8 & 1.1 & 1.4 & 2 & 0.8 & 0 \end{pmatrix}. \qquad (18)$$

In spite of the fact that the structure of cognitive map graphs represented in fig. 3a and fig . 3b is the same, weights of relations are different and cognitive maps themselves and their characteristics must not coincide. According to (18) first maximal elements to $\mathbf{K}$ – matrix - $\mathbf{K_{42}, K_{72}}$ и $\mathbf{K_{26}}$

Rank distributions $\psi$, $a\psi$, $\nu$, and $a\nu$ and their comparison with c $\psi^{imp}$, $a\psi^{imp}$, $\nu^{imp}$, and $a\nu^{imp}$, obtained with pulsed method is given in Table 4.

| $\psi$ | # |   | # | $\psi^{imp}$ | $a\psi$ | # |   | # | $a\psi^{imp}$ |
|---|---|---|---|---|---|---|---|---|---|
| 10.1  | 5 |   | 5 | 1.075  | 10.1  | 5 |   | 6 | 3.548 |
| 7.867 | 2 |   | 4 | 0.994  | 7.867 | 2 |   | 7 | 2.244 |
| 6.108 | 6 |   | 2 | 0.906  | 3.374 | 6 |   | 4 | 1.65  |
| 5.21  | 4 |   | 3 | 0.448  | 6.127 | 7 |   | 5 | 1.501 |
| 4.927 | 7 |   | 1 | 0.144  | 5.276 | 4 |   | 1 | 1.282 |
| 4.367 | 3 |   | 7 | -0.177 | 5.033 | 3 |   | 2 | 1.238 |
| 1.21  | 1 |   | 6 | -1.157 | 3.676 | 1 |   | 3 | 1.017 |

| $\nu$ | # |   | # | $\nu^{imp}$ | $a\nu$ | # |   | # | $a\nu^{imp}$ |
|---|---|---|---|---|---|---|---|---|---|
| 11.6 | 4 |   | 1 | 2.21 | 11.6 | 4 |   | 3 | 3.5  |
| 7.6  | 7 |   | 4 | 0.94 | 7.6  | 7 |   | 5 | 2.78 |
| 7.22 | 1 |   | 3 | 0.6  | 7.22 | 1 |   | 1 | 2.4  |
| 6.52 | 2 |   | 2 | 0.25 | 6.52 | 2 |   | 4 | 1.51 |
| 5.12 | 3 |   | 7 | 0.17 | 5.2  | 3 |   | 7 | 1.12 |
| 3.82 | 6 |   | 6 | -0.9 | 4.42 | 6 |   | 6 | 0.84 |
| -2.2 | 5 |   | 5 | -1.1 | 2.9  | 5 |   | 2 | 0.27 |

Table 4. Rank distribution of concepts according to the values of components $\psi$, $a\psi$, $\nu$ and $a\nu$ and $\psi^{imp}$, $a\psi^{imp}$, $\nu^{imp}$, and $a\nu^{imp}$ for the cognitive map represented in fig. 3b.

Note that with normalization $\widetilde{W}$ on different constants rank distributions of characteristics, obtained with pulsed method, will change.

## 7. DISCUSSION

Proposed in the paper K-method allows to define in cognitive map the influence of one concept on another ("pair" interaction). Note that in pulsed method that characteristic is absent. The definition of that "pair" influence, calculating of K-matrix is possible for any values of influence quantities (for any adjacency matrix of cognitive map).

Obtained matrix does not depend upon the choice of "influence concepts". Increasing of all influence quantities (adjacency matrix component) leads to the same increasing of K-matrix components. This, specifically, means that ranked series of K-matrix component values remains the same. And what's more, relations of one K-matrix components to other remain the same. Note that it is not so in pulsed method.

Learning of K-matrix components enables also to introduce and to calculate "collective" influences. For example, such a characteristic as "pressure" - ψ, which characterizes total influence of all concepts on the given. One more introduced "collective" characteristic is "consequence" - ν, the influence of the given concept on all the rest.

The introduction of K-method "collective" characteristics has made it possible to compare K-method with pulsed method. For this in pulsed method similar (but, of course, not identical) characteristics have been introduced. As appeared, such a "collective" characteristic of K-method as amplitude quantity of "pressure" and analogous to it in pulsed method gives close results.

On the whole, K-method enables to find "pair" interaction, to make these calculations for any adjacency matrix of cognitive map, to introduce characteristics of "collective" influence.

We find it is interesting to apply K-method of relations and concepts ranking for various types of complex nets.

# Appendix A

## "Pressure" and "Consequence" in pulsed method

Here in the frames of pulsed method we introduce characteristics which to some extent are analogous similar to introduced in $K$ – method pressure $\boldsymbol{\psi}$ and $\boldsymbol{a\psi}$ and consequence $\boldsymbol{v}$ and $\boldsymbol{av}$. Further we will mark these characteristics $\boldsymbol{\psi}^{imp}$, $\boldsymbol{a\psi}^{imp}$, $\boldsymbol{v}^{imp}$, and $\boldsymbol{av}^{imp}$.

At first, for convenience, write down (5) in the following way

$$\boldsymbol{v} = \boldsymbol{\Omega p}(0), \tag{A1}$$

Where $\boldsymbol{v}$ is $\boldsymbol{v}(n \to \infty)$, $\boldsymbol{\Omega} = (\boldsymbol{1} - \boldsymbol{W})^{-1}$, where as earlier, accepted $\boldsymbol{v}(init.) = 0$.

Analog of pressure on concept $\alpha$ - $\psi_\alpha$ in pulse method may serve value $\alpha$ of component of vector $\boldsymbol{\Omega p}(0)$, where vector $\boldsymbol{p}(0)$ has zero $\beta$ component and the rest are equal 1. So, for example, for cognitive map with three concepts $\psi_2^{imp}$ is defined as

$$\psi_2^{imp} = \left[\begin{pmatrix} \Omega_{11} & \Omega_{12} & \Omega_{13} \\ \Omega_{21} & \Omega_{22} & \Omega_{23} \\ \Omega_{31} & \Omega_{32} & \Omega_{33} \end{pmatrix} \begin{pmatrix} 1 \\ 0 \\ 1 \end{pmatrix}\right]_2 = \Omega_{21} + \Omega_{23} . \tag{A2}$$

That is, at the initial moment of time unity pulses on all concepts are set, except the second one, after that we calculate values $\psi_2^{imp}$ on concept 2 with $n \to \infty$. Thus, you may represent $\psi_2^{imp}$ as total action of all concepts on the second concept, see, fig. 3a.

For arbitrary component $\beta$ value $\psi_\beta^{imp}$ may be written down as

$$\psi_\beta^{imp} = \sum_k \Omega_{k\beta} - \Omega_{\beta\beta}. \tag{A3}$$

Similarly, for pulse analog of consequence $v_\alpha^{imp}$

$$v_\alpha^{imp} = \sum_k \Omega_{\alpha k} - \Omega_{\alpha\alpha}. \tag{A4}$$

Amplitude values of pulse analogs of pressure and consequence may be written down as

$$a\psi_\beta^{imp} = \sum_k |\Omega_{k\beta}| - |\Omega_{\beta\beta}|, \tag{A5}$$

$$av_\alpha^{imp} = \sum_k |\Omega_{\alpha k}| - |\Omega_{\alpha\alpha}|. \tag{A6}$$

In Table2 rank distribution of cognitive map concepts, represented in fig1a, is given. For the comparison of the obtained results beside each column of ranked $\boldsymbol{\psi}^{imp}$, $\boldsymbol{a\psi}^{imp}$, $\boldsymbol{v}^{imp}$, and $\boldsymbol{av}^{imp}$, corresponding distributions $\boldsymbol{\psi}$, $\boldsymbol{a\psi}$, $\boldsymbol{v}$ and $\boldsymbol{av}$, obtained with $K$ – method, are given.

| $\psi$ | # | # | $\psi^{imp}$ | $a\psi$ | # | # | $a\psi^{imp}$ |
|---|---|---|---|---|---|---|---|
| 6.609 | 7 | 6 | 8.265 | 6.335 | 7 | 6 | 14.056 |
| 2.985 | 6 | 9 | 3.669 | 5 | 9 | 9 | 6.622 |
| 3 | 9 | 5 | 0.874 | 4.661 | 6 | 7 | 5.505 |
| 0.885 | 5 | 3 | 0 | 3.995 | 5 | 1 | 4.342 |
| 0 | 3 | 4 | 0 | 3.816 | 1 | 5 | 2.847 |
| 0 | 4 | 8 | 0 | 2.5 | 2 | 2 | 2.5 |
| 0 | 8 | 7 | -0.519 | 0 | 3 | 3 | 0 |
| -1.304 | 1 | 1 | -1.681 | 0 | 4 | 4 | 0 |
| -2.5 | 2 | 2 | -2.5 | 0 | 8 | 8 | 0 |

| $v$ | # | # | $v^{imp}$ | $av$ | # | # | $av^{imp}$ |
|---|---|---|---|---|---|---|---|
| 8.117 | 3 | 8 | 4.552 | 8.533 | 8 | 8 | 19.273 |
| 3.66 | 1 | 3 | 4.472 | 8.177 | 3 | 3 | 5.123 |
| 0.538 | 5 | 1 | 3.309 | 3.66 | 1 | 1 | 3.96 |
| 0 | 4 | 9 | 1.309 | 3.116 | 5 | 2 | 2.849 |
| 0 | 6 | 4 | 0 | 1.158 | 2 | 5 | 2.705 |
| 0 | 7 | 6 | 0 | 1.283 | 9 | 9 | 1.96 |
| -0.281 | 8 | 7 | 0 | 0 | 4 | 4 | 0 |
| -0.843 | 9 | 2 | -2.687 | 0 | 6 | 6 | 0 |
| -3.116 | 2 | 5 | -2.849 | 0 | 7 | 7 | 0 |

Table A1. Rank distribution of concepts according to values of components $\boldsymbol{\psi}$, $\boldsymbol{a\psi}$, $\boldsymbol{v}$ and $\boldsymbol{av}$ and $\boldsymbol{\psi}^{imp}$, $\boldsymbol{a\psi}^{imp}$, $\boldsymbol{v}^{imp}$, and $\boldsymbol{av}^{imp}$ for cognitive map, represented in fig. 1.